\documentclass{elsart}

\usepackage{natbib}

\usepackage{epsfig}

\usepackage{amssymb}

\usepackage{bm}  

\begin{document}

\newcommand{\be}{\begin{equation}}
\newcommand{\ee}{\end{equation}}
\newcommand{\bea}{\begin{eqnarray}}
\newcommand{\eea}{\end{eqnarray}}
\renewcommand{\vec}[1]{\bm{#1}}

\begin{frontmatter}



\title{Models of magnetized neutron star atmospheres}


\author{V. Suleimanov}
\address{Institut f\"ur Astronomie und Astrophysik, Kepler Center for Astro and
Particle Physics, Universit\"at T\"ubingen, Sand 1, 72076 T\"ubingen, Germany;\\
Kazan State University, Kremlevskaja str., 18, Kazan 420008, Russia}
\corauth[cor]{Corresponding author}
\ead{suleimanov@astro.uni-tuebingen.de}

\author{A.Y. Potekhin}
\address{Ioffe Physical-Technical Institute,  Polytekhnicheskaya str., 26, St.
Petersburg 194021, Russia}
\author{K. Werner}
\address{Institut f\"ur Astronomie und Astrophysik, Kepler Center for Astro and
Particle Physics, Universit\"at T\"ubingen, Sand 1, 72076 T\"ubingen, Germany}

\begin{abstract}

We present a new computer code for modeling magnetized neutron star atmospheres
in a wide range of magnetic fields ($10^{12} - 10^{15}$ G) and
effective temperatures ($3 \times 10^5 - 10^7$ K).
The atmosphere is assumed to consist either of
fully ionized electron-ion plasmas or
of partially ionized hydrogen.
Vacuum resonance and partial mode conversion are taken into account.
Any inclination of the magnetic field relative
to the stellar surface is allowed.
We use modern opacities of fully or partially ionized plasmas
in strong magnetic fields and solve the coupled radiative transfer equations
for the normal electromagnetic modes in the plasma.
Using this code, we study the possibilities to explain the soft X-ray
spectra of isolated neutron stars
by different atmosphere models.
 In particular, the outgoing spectrum
using the ``sandwich'' model (thin atmosphere with a hydrogen layer above a helium
layer) is constructed.
Thin partially ionized hydrogen atmospheres with
vacuum polarization are shown
to be able to improve our understanding of
  the observed spectrum
of the nearby isolated neutron star RBS 1223 (RX J1308.8+2127).

\end{abstract}


\end{frontmatter}

\parindent=0.5 cm

\section{Introduction}

At present several classes of neutron stars (NSs) with strong magnetic field
are known. They include X-ray dim isolated NSs
(XDINSs, \citealt{Haberl:07}), 
central compact objects (CCOs) in supernova 
remnants \citep{Pavlovetal:04}, anomalous X-ray pulsars 
and soft-gamma repeaters (AXPs and SGRs; see reviews by 
\citealt{Kaspi:07,Mereghettietal:07,Mereghetti:08}). 
The NSs in these 
classes have superstrong magnetic fields up to $B \ge 10^{14}$ G (SGR and AXP) and 
$B \sim \mbox{a few} \times 10^{13}$~G in XDINSs, 
evaluated from period changes and from absorption features in the observed 
spectra, if they are interpreted as ion cyclotron lines (see reviews by 
\citealt{Haberl:07,vKK:07}). 
 
These NSs  are 
sufficiently hot ($T_{\rm eff} \sim 10^6 - 10^7$ K) to be observed as thermal 
soft X-ray sources.
Some of the XDINSs and CCOs  have one or more absorption features in their X-ray spectrum at the 
energies 0.2 -- 0.8 keV \citep{Haberl:07}, and 
 the central energies of these features appear to be 
harmonically spaced \citep{Sanwaletal:02,big:03,Swopeetal:07,vKK:07,Haberl:07}.
The optical/ultraviolet fluxes of the known XDIN optical 
counterparts are a few 
times larger than the  
blackbody extrapolation of the X-ray spectra 
\citep{Burwitzetal:01,Burwitzetal:03,Kaplanetal:03,Motchetal:03,Mignanietal:07}. 
 
The XDINs are nearby objects, and parallaxes of some of them have 
been measured \citep{Kaplanetal:02a}. Therefore, they give a good possibility to 
measure the NS 
radii, yielding useful information on the equation of state 
(EOS) for the NS inner core \citep{Trumperetal:04,LP07}. 
For a sufficiently accurate evaluation of NS radii, 
a good model of the NS surface radiation is necessary for 
the observed X-ray spectra fitting.

Structures and emergent spectra of  NS 
atmospheres with strong ($B \ge 10^{12}$ G) magnetic fields 
have been modeled by many scientific 
groups 
\citep{Shibanovetal:92, Rajagopaletal:97, Ozel:01, Ho.Lai:01, Ho.Lai:03, 
Ho.Lai:04,vAL:06}. 
Methods of fully 
ionized model atmospheres  modeling (see, e.g., \citealt{Zavlin:09} 
for references) and partially ionized hydrogen 
atmospheres modeling 
\citep{Potekhinetal:04, Ho.Lai:04, Hoetal:08} were developed. 
Mid-$Z$ element atmospheres for strongly 
magnetized NSs have also been modeled  
\citep{MH:07}.
The effect of the vacuum polarization on magnetized NS atmospheres was 
studied by \citet{PG:84,Lai.Ho:02, Lai.Ho:03}, and 
model atmospheres 
with partial mode conversion due to vacuum 
polarization have been computed \citep{Ho.Lai:03,vAL:06}. 
 Some of the XDINSs 
 might have a ``thin'' atmosphere above the condensed surface, which 
could be optically thick to low-energy photons and optically 
thin to high-energy photons
\citep{Motchetal:03,Hoetal:07}. 

 Here we present a new computer code,  which allows to compute magnetized 
NS model atmospheres consisting of partially ionized hydrogen, taking into account the vacuum
polarization effect together with partial mode conversion and an arbitrary magnetic field 
inclination. Compared to previous codes, other 
methods were used for the temperature correction procedure and the solution of the
radiation transfer equation. We also present some new results, which were obtained by using this
code.
 
\section{Method of atmosphere structure calculations} 
\label{s:methods} 

We computed model
atmospheres of hot, magnetized NSs subject to the constraints 
of hydrostatic and radiative equilibrium assuming planar geometry.
There are two versions of the code:
the first one is designed for the magnetic field 
$\vec{B}$ perpendicular to the surface, and in  the second version, the angle 
$\theta_B$ between $\vec{B}$ and $\vec{n}$ is arbitrary, 
and calculations are more expensive.

The model atmosphere structure for an NS with effective temperature $T_{\rm
eff}$, surface gravity $g$, magnetic field $B$, and given chemical composition is described by 
a set of
equations  (e.g., \citealt{Shibanovetal:92,Ho.Lai:01} and references
therein), which are written below for
the simplest first case, where $\theta_B=0$. 
In the second case (for arbitrary $\theta_B$, e.g., \citealt{Zavlin:95}), there is
an additional dependence of the absorption ($k_{\nu}^i$)
and scattering ($\sigma_{\nu}^i$) coefficients
on the azimuthal angle
$\varphi$; accordingly, there are additional integrations over $\varphi$
in Eqs.~(\ref{g_rad}), (\ref{sf}), (\ref{econs}),
and (\ref{e:eta}) in the second case.

The hydrostatic equilibrium
equation reads
\be \label{e:hyd}
  \frac {d P_{\rm g}}{dm} = g - g_{\rm rad},
\ee
where 
\be
    g=\frac{GM_{\rm NS}}{R^2_{\rm NS}\sqrt{1-R_{\rm S}/R_{\rm NS}}},
\ee
\be \label{g_rad}
 g_{\rm rad} =  \frac{2 \pi}{c} \sum_{i=1}^2 \int_0^{\infty}\, d\nu \, 
\int_{-1}^{+1} (k_{\nu}^i+\sigma_{\nu}^i) \, \mu \, I_{\nu}^i(\mu) \, d\mu,
\ee
and $R_{\rm S}=2GM_{\rm NS}/c^2$ is the Schwarzschild radius.
 $I_{\nu}^i(\mu)$ is the
specific intensity in mode $i$,  $\mu = \cos\theta$, where $\theta$ is the angle between 
the surface normal and the radiation propagation direction,
 $P_{\rm g}$ is  the  gas pressure, and the column density $m$ is
determined as

\be
     dm = -\rho \, dz \, .
\ee
The variable $\rho$ denotes the gas density and $z$ is the vertical distance.

The radiation transfer equations for the two modes are
\be \label{rtr}
   \mu\frac{d I_{\nu}^i}{d \tau_{\nu}^i} =  I_{\nu}^i - S_{\nu}^i
\ee
where the source function is
\be \label{sf}
S_{\nu}^i = \frac{k_{\nu}^i}{k_{\nu}^i+\sigma_{\nu}^i} \frac{B_{\nu}}{2}
 + \, \frac{1}{k_{\nu}^i+\sigma_{\nu}^i} \sum_{j=1}^2 \, \int_{-1}^{+1} \sigma_{\nu}^{ij}(\mu,\mu')\, I_{\nu}^j (\mu')\, d\mu' \, , 
\ee
and $B_{\nu}$  is the blackbody (Planck)
intensity.
 The optical
depth $\tau_{\nu}^i$ is defined as
\be
    d \tau_{\nu}^i = (k_{\nu}^i+\sigma_{\nu}^i) \, dm.
\ee
Of course, opacities $k_{\nu}^i$ and $\sigma_{\nu}^i$ depend on $\mu$.
These equations have to be completed by the energy balance equation
\be  \label{econs}
  \int_0^{\infty} \, d\nu  \sum_{i=1}^2 \, \int_{-1}^{+1} \left( (k_{\nu}^i +\sigma_{\nu}^i)  I_{\nu}^i(\mu) 
- \eta_{\nu}^i(\mu) \right) \, d\mu = 0  
\ee
with emissivity
\be \label{e:eta}
\eta_{\nu}^i(\mu) =  k_{\nu}^i \frac{B_{\nu}}{2} +  \sum_{j=1}^2 \int_{-1}^{+1} 
\sigma_{\nu}^{ij}(\mu,\mu')\, I_{\nu}^j (\mu') \, d\mu',
\ee
  
Equations (\ref{e:hyd}) -- (\ref{e:eta}) 
must be completed by the EOS and the charge 
and particle conservation laws. In the code two different cases of these 
 laws are considered. 
 
In the first (simplest) case, 
 a fully ionized atmosphere 
is calculated. Therefore, the EOS is 
the ideal gas law 
\be   \label{gstat} 
    P_{\rm g} = n_{\rm tot} kT \,, 
\ee 
where $n_{\rm tot}$ is the number density of all particles. 
Opacities are calculated in the same way as in the paper by 
\cite{vAL:06} (see references therein for the background theory 
and more sophisticated approaches). 
The vacuum polarization effect is taken into account 
following the same work. 
 
The second considered case is a partially ionized hydrogen atmosphere. 
In this case the EOS and the corresponding 
opacities are taken from tables calculated by 
\citet{PCh:03,PCh:04}. The normal mode polarization vectors 
are taken from the calculations by 
 \cite{Potekhinetal:04}. The vacuum polarization effect 
is also included. 
  
 We use a
logarithmically equidistant photon energy set in our computations in the 
range 0.001 -- 20 keV with additional points near each ion 
cyclotron resonance. 
If the vacuum resonance is taken into 
consideration, then 
another photon energy grid is used, which is constructed using the ``equal 
grid'' method \citep{Ho.Lai:03}. 
This energy grid is recalculated after every iteration.

The radiation transfer 
equation (\ref{rtr}) is  solved on a set of 40 polar angles $\theta$
(and additionally 6 azimuthal angles $\varphi$ in the case of inclined magnetic field) 
by the short characteristic method \citep{Ols.Kun:87}. 
We use the conventional condition (no external radiation) at the outer boundary. 
 At the inner boundary we take the incoming specific intensities equal to blackbody
radiation corresponding to the temperature of the last atmosphere point.
In the case of the thin atmosphere the temperature at the inner boundary is considered as 
the temperature of a condensed NS surface. 
 The code allows one to take into account the partial mode conversion 
according to \cite{vAL:06}.
 
The solution of the radiative transfer equation (\ref{rtr}) is checked 
for the energy balance equation (\ref{econs}).
Temperature corrections are then evaluated using three different 
procedures  \citep{Kurucz:70},
modified to deal with  strong magnetic fields. 
A nonmagnetic version of this  code was described, for example, by
\citet{Ibragimov.etal:03} and \citet{Sul.Wer:07}.

The iteration procedure is repeated until the relative flux error is 
smaller than 1\% and the relative flux derivative error is smaller 
than 0.01\%.
 
Our method of calculation has been tested by a comparison to models for magnetized NS 
atmospheres \citep{Shibanovetal:92, Pavlovetal:94, Ho.Lai:01, Ozel:01, 
Ho.Lai:03}.  Model atmospheres with partially ionized 
hydrogen are compared to models computed by \cite{Hoetal:07}.
We have found that our models are 
in a good agreement with these  calculations. 
 
\section {Results} 
\label{s:results} 
 
Here we present some new results, obtained by using of the developed code.
 In all calculations below we use 
the same surface gravity, $\log g$ = 14.3. 

\begin{figure} 
\includegraphics[width=0.9\columnwidth]{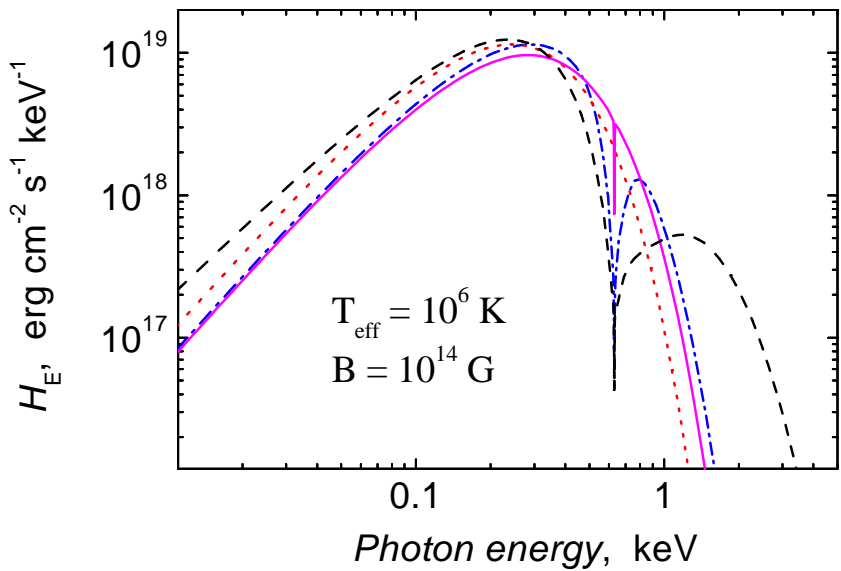} 
\caption{\label{f:fig4a} 
Emergent spectra of the thin fully ionized hydrogen atmospheres above a 
 solid surface with $T_{\rm eff} = 10^6$ K and $B = 10^{14}$ G in 
 comparison to the semi-infinite atmosphere (dashed curve). The 
 spectra of atmospheres with  surface densities $\Sigma$=1 (solid 
curve) and 100 g cm$^{-2}$ (dash-dotted curve) together with the corresponding 
blackbody (dotted curve) are shown. 
} 
\end{figure}

One of the problems related to AXPs 
is the lack of any absorption feature at the 
proton cyclotron energy. \citet{Ho.Lai:03} 
suggested that a possible 
solution of this problem is the 
suppression of the cyclotron absorption feature due to the 
vacuum polarization.
We 
demonstrate that this 
absorption line is also further reduced in a thin atmosphere. 
Figure~\ref{f:fig4a} demonstrates emergent spectra of 
 thin atmospheres without allowance for 
the vacuum polarization effect. 
The absorption 
feature disappears with decreasing 
the atmosphere surface density $\Sigma$. Equivalent widths of the absorption feature are
$\approx$ 670 eV for the semi-infinite atmosphere, $\approx$ 190 eV for the $\Sigma$ = 100 g cm$^{-2}$ 
slab and $\approx$ 5 eV for the $\Sigma$ = 1 g cm$^{-2}$ slab. 
This decrease
of the equivalent width corresponds to the decrease of the
optical depth according to the curve of growth. 
 
\begin{figure} 
\includegraphics[width=0.9\columnwidth]{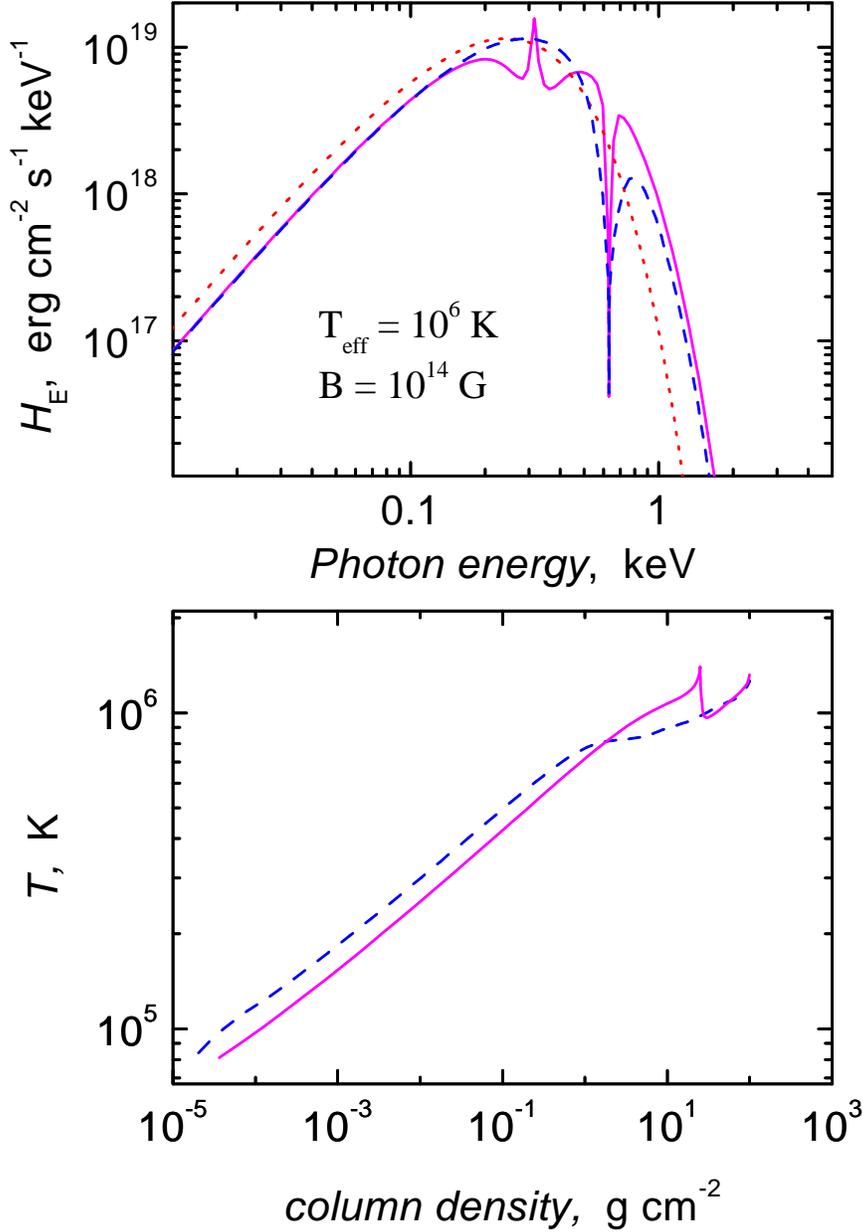} 
\caption{\label{f:fig4b} 
 Emergent spectrum (top panel)  and temperature structure (bottom panel)
 of the ``sandwich'' model atmosphere above a 
 solid surface with $T_{\rm eff} = 10^6$ K and $B = 10^{14}$ G (solid 
curve) in comparison with the thin fully ionized hydrogen atmosphere 
 (dashed curve) with the same parameters.  The surface densities of both 
 model atmospheres are 100 g cm$^{-2}$, in the ``sandwich'' model the H slab 
has 25 g cm$^{-2}$ surface density and the He slab has  75 g cm$^{-2}$. The 
 corresponding blackbody spectrum (dotted curve) is also shown in the top
panel. } 
\end{figure} 

Various hypotheses  were considered for an explanation  
of the two harmonically spaced absorption features in CCOs
(see \citealt{MH:07} and references therein). 
Here we 
suggest another one, which we name ``sandwich atmosphere''. 
A thin, chemically layered atmosphere above a condensed NS surface can arise from 
accretion of interstellar gas with cosmic chemical composition. In 
this case, hydrogen and helium quickly 
separate due to the high  
gravity. 
In this ``sandwich atmosphere'' a layer 
of hydrogen is located above a helium slab, 
and the emergent spectrum has two 
absorption features, corresponding to proton and $\alpha$-particle 
cyclotron energies. In Fig.~\ref{f:fig4b}  the emergent spectrum for one of 
such models  together with the corresponding temperature structure are shown. 
Fully ionized hydrogen and helium layers are considered in this model. 
The emission feature at the helium absorption line 
arises due to a local temperature bump at the boundary between the helium and 
hydrogen layers. This temperature increase compensates a lower number density in the hydrogen
layer to avoid a gas pressure jump. Of course, this temperature jump should be smoothed by 
 hydrogen-helium mixing and  thermal conduction flux at the boundary between layers.
Here we used the sharp boundary and did not take into
account thermal conductivity. Results can also  change quantitatively, if partially ionized
hydrogen and helium layers are considered.
 
\begin{figure} 
\includegraphics[width=0.9\columnwidth]{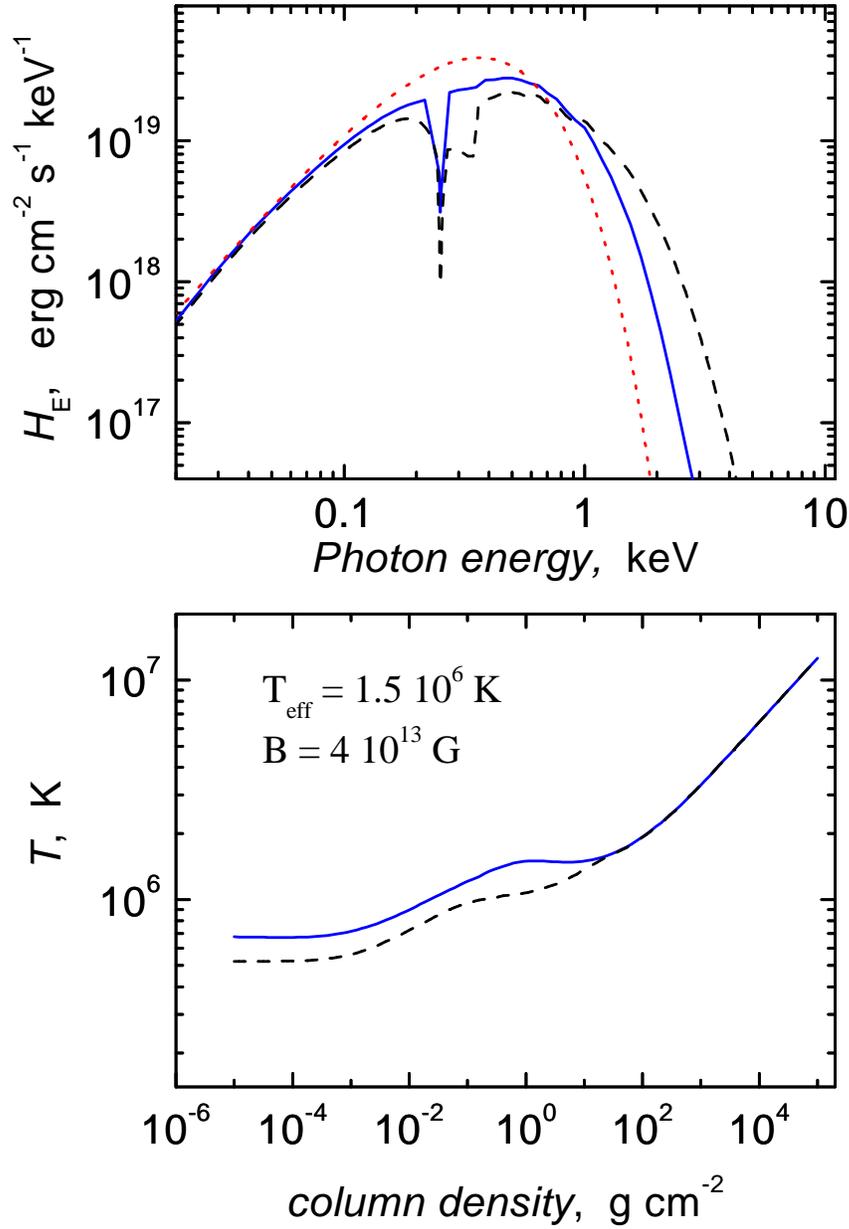} 
\caption{\label{f:fig8} 
Emergent spectra and temperature structures of the partially ionized 
 hydrogen model atmospheres  with $T_{\rm eff} = 1.5 \times 
10^6$ K with (solid curves) and without (dashed curves) vacuum polarization 
effect (with the partial mode conversion). The magnetic field strength is $B=4 \times 
10^{13}$ G. The corresponding blackbody spectrum is also shown in the 
upper panel (dotted curve).  
} 
\end{figure} 
 
\begin{figure} 
\includegraphics[width=0.9\columnwidth]{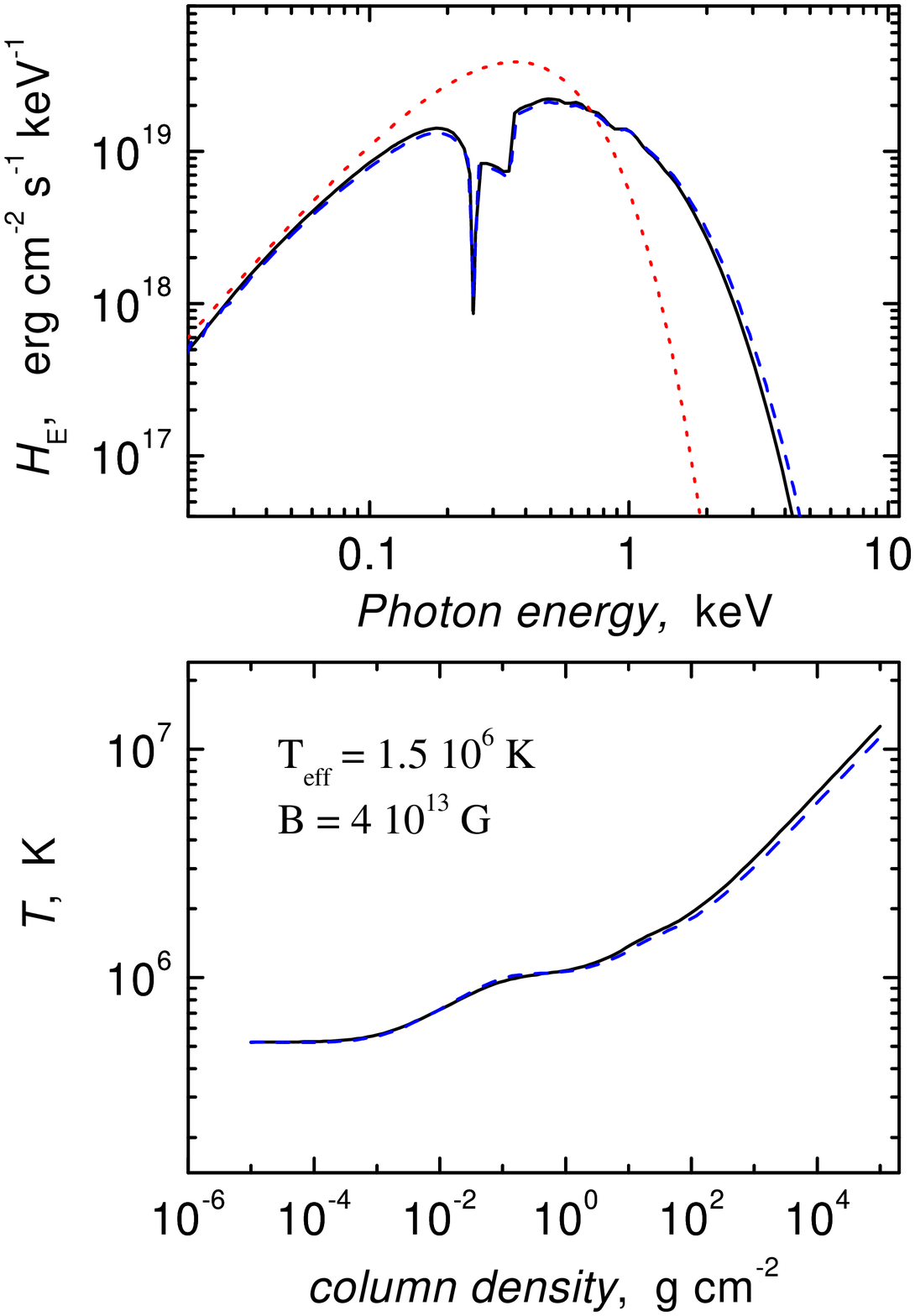} 
\caption{\label{f:fig7} 
Emergent spectra and temperature structures of partially ionized 
 hydrogen model atmospheres  with $T_{\rm eff} = 1.5 \times 
10^6$ K with  inclinations of the magnetic field ($B=4 \times 
10^{13}$ G) to the surface normal $\theta_B$ equal 0$^{\circ}$ (solid 
curves) and 90$^{\circ}$ (dashed curves).  The vacuum polarization effect is not 
included.  The corresponding blackbody spectrum  is also shown in 
the upper panel (dotted curve).  
} 
\end{figure}

Most of the XDINSs have magnetic fields $B \ge 
10^{13}$ G and  color temperatures $\approx 10^6$~K \citep{Haberl:07}. 
Hydrogen model atmospheres are partially ionized under these conditions 
and the vacuum polarization effect is also significant. 
Here we present  first results of modeling of partially ionized 
hydrogen atmospheres with the vacuum polarization effect and partial mode 
conversion using our radiative transfer code. 
 Previously
 partial mode conversion  was considered only for 
fully ionized atmosphere models \citep{vAL:06}.
In 
Fig.~\ref{f:fig8} we compare  spectra and temperature 
structures of the partially ionized hydrogen model atmospheres with and 
without the partial mode conversion 
effect. 
When the X-mode (having smaller opacity) partially converts to the 
O-mode in the surface layers of the atmosphere, the energy absorbed by 
the O-mode heats these upper layers. As a 
result, the emergent spectra are closer to the blackbody.  
The absorption feature to the
right of the proton cyclotron line arises mainly due to the
transition from the ground state to the first excited state of
the H atom; it has a maximum at 350 eV and is strongly broadened
at lower energies because of the atomic motion effect
(cf.\ Fig.~5 in \citealt{PCh:04} and related discussion there).
 
\begin{figure} 
\includegraphics[width=0.9\columnwidth]{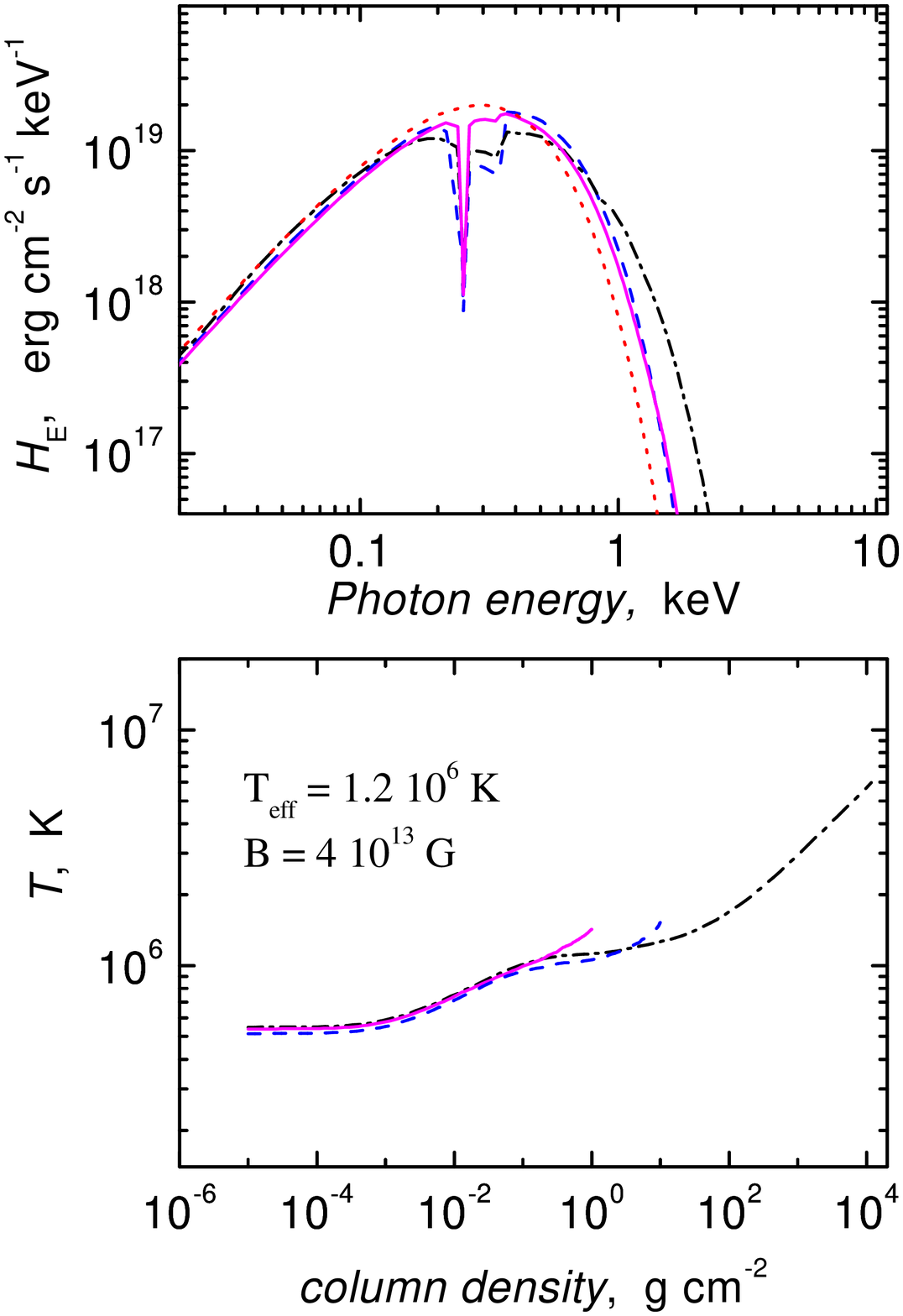} 
\caption{\label{f:fig9} 
Emergent spectra and temperature structures of the partially ionized 
 hydrogen model atmospheres with $T_{\rm eff} = 1.2 \times 
10^6$ K with vacuum polarization 
effect  and partial mode conversion shown for various surface densities $\Sigma$ 
(solid curves -- 1 g cm$^{-2}$, dashed curves -- 10 g cm$^{-2}$, 
 dash-dotted curves -- semi-infinite 
atmosphere).  The magnetic field strength is $B=4 \times 10^{13}$ G.  The corresponding 
blackbody spectrum  is also 
shown in the upper panel (dotted curve). 
} 
\end{figure} 

{\rm The effective 
temperature and magnetic field strengths are not uniform 
over the NS surface, and generally the magnetic field is 
not perpendicular to the surface. Therefore, for comparison to observations, 
it will be necessary to integrate the local 
model spectra over the NS surface (see \citealt{Zavlin:95,Hoetal:08}), and 
 to compute model atmospheres with inclined magnetic 
field. This possibility is included in our code. For example, 
Fig.~\ref{f:fig7} shows 
spectra and temperature structures of partially ionized hydrogen model atmospheres 
with magnetic field  perpendicular and  parallel to the NS surface. The vacuum polarization
effect is taken into account without mode conversion. The difference between emergent spectra is not
significant (see also the first models with different magnetic field inclinations in 
\citealt{Kaminkeretal:82,Shibanovetal:92}), but it can be more significant at other 
atmospheric parameters \citep{Hoetal:08}.}

\cite{Hoetal:07}  demonstrated that a single partially ionized thin 
hydrogen atmosphere can explain the problem of the observed optical/UV flux excess
over the blackbody extrapolation from X-ray to the optical range  
in the case of brightest 
isolated NS RX\,J1856.4$-$3754: the model fits well both 
the observed optical flux and the X-ray spectrum. 
RX\,J1856.4$-$3754 has very low pulsed fraction of radiation 
($\approx 1.2$\%, 
\citealt{Tie.Meregh:07}), therefore it is possible 
to fit the radiation of this star by a 
single model atmosphere (although the small pulsed fraction can
also be explained by a small value of an angle between the
magnetic and rotation axes or between the rotation axis and the
line of sight).
 Other XDINSs have larger pulse 
fractions, up to 18\% (RBS 1223, \citealt{Haberletal:04}). 
In this case the temperature  is certainly not uniform across the NS surface, 
and the 
excess optical flux can be explained by the radiation from cool surface parts 
\citep{Swopeetal:05}. 
 
\begin{figure} 
\includegraphics[width=0.9\columnwidth]{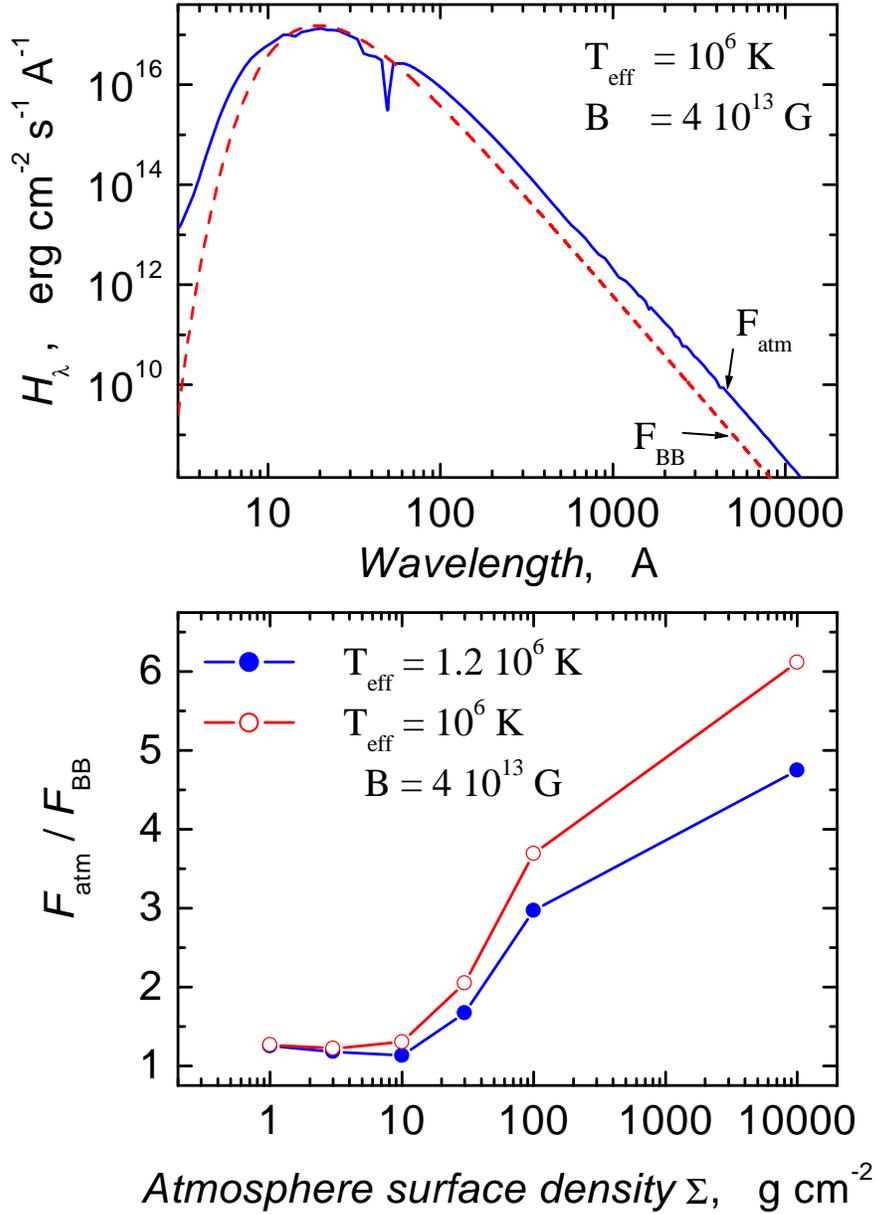} 
\caption{\label{f:fig10} 
{\it Top panel:} Emergent spectrum of the partially ionized hydrogen model 
atmosphere of neutron stars with $T_{\rm eff} = 
10^6$ K with vacuum polarization 
effect,  partial mode conversion and  magnetic field strength $B=4 \times 
10^{13}$ G.  The blackbody spectrum fitted to  the maximum  of the 
spectral distribution is also shown (dashed curve). At the optical band the model 
atmosphere flux is a few times larger than the blackbody flux. 
{\it Bottom panel:} Ratios of the model atmosphere flux to the blackbody 
(X-ray fitted) flux at the optical band depending on the model atmosphere thickness 
(surface density $\Sigma$) for the models with $T_{\rm eff} = 10^6$ K  and 
$T_{\rm eff} = 1.2 \times 10^6$ K. 
} 
\end{figure} 
 
We now investigate properties of partially ionized hydrogen models,
 which can be applied to the RBS 1223 atmosphere. The color 
temperature of this star,  found from  X-ray spectra fitting, 
is close 
to $10^6$ K, with magnetic field $B\approx 4 \times 10^{13}$ G 
\citep{Swopeetal:07}.  In particular, we study the optical flux excess 
in comparison 
to the X-ray fitted blackbody flux in this kind of models in order 
to explore which part of the observed optical excess can be explained by 
the atmosphere effect. This optical excess is illustrated
in the top panel of Fig.~\ref{f:fig10}. 
 For this aim we have calculated two sets of  
 models with vacuum polarization and partial mode 
conversion. The models in the first set have 
effective temperatures $T_{\rm eff} = 10^6$ K and the models of second one 
have effective temperatures $T_{\rm eff} = 1.2 \times 10^6$ K. In both 
sets $B=4 \times 10^{13}$ G, and models with 
surface densities $\Sigma$ = 1, 3, 10, 30, 100 and 10$^5$ (semi-infinite 
model) g cm$^{-2}$ are computed.  In Fig.~\ref{f:fig9} we show  emergent spectra 
and  temperature structures for some models from the second set. 
Clearly, the X-ray spectra of the models with $\Sigma \le 
10$ g cm$^{-2}$ are close to a blackbody and, therefore, better fit 
the observed X-ray spectrum.  Let
us remark that there is a significant absorption feature in the
RBS 1223 spectrum \citep{Swopeetal:07}, which cannot be fitted by
the simple blackbody spectrum. Here we do not try to fit the
observed spectrum of RBS 1223 by the single temperature model:
clearly, such fitting would require the integration of the local
spectra over the neutron star surface assuming some non-uniform
temperature and a magnetic field distributions (see
\citealt{Zavlin:95,Hoetal:08}). Instead, we are trying to
understand, what kind of atmosphere (thin or  semi-infinite) 
be more appropriate for this modeling. 
 
 In Fig.~\ref{f:fig10} (bottom panel) we show 
the ratio of the model 
atmosphere flux to the X-ray fitted  (in the 0.4 - 1 keV range) blackbody flux 
in the optical band  at $\lambda$ = 5150 \AA depending on $\Sigma$ for 
both sets. 
The observed ratio is about 5 \citep{Kaplanetal:02b}, in
agreement with the semi-infinite atmosphere models. However, the
observed blackbody like X-ray spectrum agrees with the thin
atmosphere models, for which this ratio is close to 1. Therefore,
the observed optical excess cannot be explained by the thin
atmosphere model alone; instead, it can arise due to a nonuniform
surface temperature distribution, in agreement with the RBS 1223
light curve modeling  \citep{Swopeetal:05}.

 
\section{Conclusions} 
\label{s:conclusions} 
 
In this work we present a new code, which can model fully ionized 
 and  partially ionized hydrogen atmospheres in a wide range 
of effective temperatures ($3 \times 10^5$ -- 10$^7$ K) and magnetic 
fields (10$^{12}$ -- 10$^{15}$ G), with any inclination 
of the magnetic 
field to the stellar surface.  The vacuum polarization effect with 
partial mode conversion is taken into consideration. Calculated 
emergent spectra and temperature structures of the model 
atmospheres agree with previously published ones. 
 
We have studied the 
properties of  thin atmospheres above condensed NS surfaces. We 
demonstrated that the proton cyclotron absorption line disappears in the 
thin hydrogen model atmospheres. A new thin ``sandwich'' model atmosphere 
(hydrogen layer above helium layer)  is proposed to explain the occurrence 
of two absorption features in the observed X-ray spectra of some isolated NSs. 

A set of of
partially ionized hydrogen model atmospheres with vacuum polarization
and partial mode conversion  with  parameters  (effective 
temperature and the magnetic field strength) close to the probable parameters 
of the isolated NS RBS\,1223  were calculated. 
We analysed the optical excess (relative to
the X-ray fitted blackbody flux) in the model spectra and 
found that the optical flux excess $\approx 5$ for the semi-infinite model 
atmospheres decreases down to 1 with decreasing surface density $\Sigma$ 
of the atmosphere. Spectra of thin model atmospheres are closer to the observed 
RBS\,1223 X-ray spectrum, therefore we conclude that the observed optical 
excess should be explained by nonuniform surface temperature 
distribution. 
 
{\rm Acknowledgements} VS thanks DFG for financial support (grant We 1312/35-1 and grant 
SFB/Transregio 7 "Gravitational Wave Astronomy") and the President's 
programme for support of leading science schools (grant NSh-4224.2008.2). 
The work of AYP is supported by RFBR grant 08-02-00837 
 and the President's 
programme for support of leading science schools (grant NSh-2600.2008.2).


\begin{thebibliography}{}


\bibitem[{{Bignami} {et~al.}(2003){Bignami},  {Caraveo}, {De Luca},  \& {Mereghetti}}] 
{big:03} 
{Bignami}, G. F., {Caraveo}, P. A., {De Luca}, A. \& {Mereghetti}, S.
The magnetic field of an isolated neutron star from X-ray cyclotron absorption lines.
 Nature 423, 725-727, 2003. 

\bibitem[{{Burwitz} {et~al.}(2001){Burwitz}, {Zavlin}, {Neuh\"auser} {et~al.}}] 
{Burwitzetal:01} {Burwitz}, V., {Zavlin}, V.E., {Neuh\"auser}, R. et~al. 
The Chandra LETGS high resolution X-ray spectrum of the isolated neutron star RX J1856.5-3754.
 A\&A 379, L35-L38, 2001. 
 
\bibitem[{{Burwitz} {et~al.}(2003){Burwitz}, {Haberl}, {Neuh\"auser} {et~al.}}] 
{Burwitzetal:03} {Burwitz}, V., {Haberl}, F., {Neuh\"auser}, R. et~al. 
The thermal radiation of the isolated neutron star RX J1856.5-3754 observed with 
Chandra and XMM-Newton. A\&A 399, 1109-1114, 2003.
 
\bibitem[{{Haberl} {et~al.}(2004) {Haberl}, {Motch}, {Zavlin} {et~al.}}] 
{Haberletal:04} 
{Haberl}, F., {Motch}, C., {Zavlin}, V.E. {et~al.} 
The isolated neutron star X-ray pulsars RX J0420.0-5022 and RX J0806.4-4123:
 New X-ray and optical observations. A\&A 424, 635-645, 2004.
 
\bibitem[{{Haberl}(2007)}]{Haberl:07} 
{Haberl}, F. 
The magnificent seven: magnetic fields and surface temperature distributions. 
A\&SS 308, 181-190, 2007.
 
\bibitem[{{Ho} \& {Lai}(2001)}]{Ho.Lai:01} 
{Ho}, W.~C.~G. \& {Lai}, D.
Atmospheres and spectra of strongly magnetized neutron stars.
 MNRAS 327, 1081-1096, 2001. 
 
\bibitem[{{Ho} \& {Lai}(2003)}]{Ho.Lai:03} 
{Ho}, W.~C.~G. \& {Lai}, D. 
Atmospheres and spectra of strongly magnetized neutron stars - 
II. The effect of vacuum polarization. MNRAS. 338, 233-252, 2003.
 
\bibitem[{{Ho} \& {Lai}(2004)}]{Ho.Lai:04} 
{Ho}, W.~C.~G. \& {Lai}, D. 
Spectral Features in the Thermal Emission from Isolated Neutron Stars: Dependence on 
Magnetic Field Strengths.  ApJ 607, 420-425, 2004.
 
\bibitem[{{Ho} {et~al.}(2007){Ho}, {Kaplan}, {Chang}, {van Adelsberg} 
\& {Potekhin}}]{Hoetal:07} 
{Ho}, W.~C.~G.,  {Kaplan}, D.L., {Chang}, P., {van Adelsberg}, M. \& {Potekhin}, 
 A.Y. 
Magnetic hydrogen atmosphere models and the neutron star RX J1856.5-3754.
 MNRAS 375, 821-830, 2007. 
 
\bibitem[{{Ho} {et~al.}(2008){Ho}, {Potekhin} \& {Chabrier}}]{Hoetal:08} 
{Ho}, W.~C.~G., {Potekhin}, A.Y. \& {Chabrier}, G. 
Model X-Ray Spectra of Magnetic Neutron Stars with Hydrogen Atmospheres.
 ApJ Suppl. Ser. 178, 102-109, 2008.
 
\bibitem[{{Ibragimov} {et~al.}(2003){Ibragimov}, {Suleimanov}, {Vikhlinin}, \& 
  {Sakhibullin}}]{Ibragimov.etal:03} 
{Ibragimov}, A.~A., {Suleimanov}, V.~F., {Vikhlinin}, A., \& {Sakhibullin}, N.~A. 
Supersoft X-ray Sources. Parameters of Stellar Atmospheres.  Astronomy Rep. 47, 186-196, 2003. 
 
\bibitem[{{Kaminker} {et~al.}(1982){Kaminker}, {Pavlov}, {Shibanov}}] 
{Kaminkeretal:82} {Kaminker}, A.D., {Pavlov}, G.G., {Shibanov}, Yu.A.
Radiation for a strongly-magnetized plasma - The case of predominant scattering.
Astroph. Space Sci. 86, 249-297, 1982.  

\bibitem[{{Kaplan} {et~al.}(2002a){Kaplan}, {van Kerkwijk}, {Anderson}}] 
{Kaplanetal:02a} {Kaplan}, D.L., {van Kerkwijk}, M.H. {Anderson}, J. 
The Parallax and Proper Motion of RX J1856.5-3754 Revisited. ApJ 571, 447-457, 2002a.  

\bibitem[{{Kaplan} {et~al.}(2002b){Kaplan}, {Kulkarni} \& {van Kerkwijk}}] 
{Kaplanetal:02b} {Kaplan}, D.L.,  {Kulkarni}, S.R. \& {van Kerkwijk}, M.H. 
A Probable Optical Counterpart to the Isolated Neutron Star RX J1308.6+2127.
 ApJ 579, L29-L32, 2002b.
 
\bibitem[{{Kaplan} {et~al.}(2003){Kaplan}, {van Kerkwijk}, {Marshall} {et~al.}}] 
{Kaplanetal:03} {Kaplan}, D.L., {van Kerkwijk}, M.H., {Marshall}, H.L., et al. 
The Nearby Neutron Star RX J0720.4-3125 from Radio to X-Rays. ApJ 590, 1008-1019, 2003. 
 
\bibitem[{{Kaspi}(2007)}]{Kaspi:07} 
{Kaspi}, V.M. Recent progress on anomalous X-ray pulsars. A\&SS 308, 1-11, 2007.
 
\bibitem[{{Kurucz}(1970)}]{Kurucz:70} 
{Kurucz}, R.~L. Atlas: a Computer Program for Calculating Model Stellar Atmospheres.
 SAO Special Report 309, 1970. 
 
\bibitem[{{Lai} \& {Ho}(2002)}]{Lai.Ho:02} 
{Lai}, D. \& Ho. W.C.G. 
Resonant Conversion of Photon Modes Due to Vacuum Polarization in a Magnetized Plasma:
 Implications for X-Ray Emission from Magnetars. ApJ 566, 373-377, 2002.
 
\bibitem[{{Lai} \& {Ho}(2003)}]{Lai.Ho:03} 
{Lai}, D. \& Ho. W.C.G. 
Transfer of Polarized Radiation in Strongly Magnetized Plasmas and Thermal Emission 
from Magnetars: Effect of Vacuum Polarization. ApJ 588, 962-974, 2003. 
 
\bibitem[{{Lattimer} \& {Prakash}(2007)}]{LP07} 
{Lattimer}, J.M. \& {Prakash}, M. 
Neutron star observations: Prognosis for equation of state constraints.
Phys. Rep. 442, 109-165, 2007. 
 
\bibitem[{{Mereghetti}(2008)}]{Mereghetti:08} 
{Mereghetti}, S. The strongest cosmic magnets: soft gamma-ray repeaters and anomalous X-ray pulsars.
A\&A Rev. 15, 225-287, 2008.
 
\bibitem[{{Mereghetti} {et~al.}(2007){Mereghetti}, {Esposito} 
\& {Tiengo}}]{Mereghettietal:07} 
{Mereghetti}, S., {Esposito}, P. \& {Tiengo}, A.
XMM Newton observations of soft gamma-ray repeaters. A\&SS 308, 13-23, 2007. 
 
\bibitem[{{Mignani} {et~al.}(2007){Mignani}, {Bagnulo}, 
{De Luca}{et~al.}}]{Mignanietal:07} 
{Mignani}, R.P,  {Bagnulo}, S., {De Luca}, A. et~al. 
Studies of neutron stars at optical/IR wavelengths. A\&SS 308, 203-210, 2007. 
 
\bibitem[{{Mori} \& {Ho}(2007)}] 
{MH:07} {Mori}, K. \& {Ho}, W.C.G. 
Modelling mid-Z element atmospheres for strongly magnetized neutron stars. 
MNRAS 377, 905-919, 2007. 
 
\bibitem[{{Motch} {et~al.}(2003) {Motch}, {Zavlin} \& {Haberl}}] 
{Motchetal:03} {Motch}, C., {Zavlin}, V.E. \& {Haberl}, F. 
The proper motion and energy distribution of the isolated neutron star RX J0720.4-3125.
A\&A 408, 323-330, 2003.
 
\bibitem[{{Olson} \& {Kunasz}(1987)}] 
{Ols.Kun:87} {Olson}, G.L. \& {Kunasz}, P.B. 
Short characteristic solution of the non-LTE transfer problem by operator perturbation.
 I. The one-dimensional planar slab. JQSRT 38, 325-336, 1987. 
 
\bibitem[{{\"Ozel}(2001)}]{Ozel:01} 
{\"Ozel}, F. 
Surface Emission Properties of Strongly Magnetic Neutron Stars. ApJ 563, 276-288, 2001.
 
\bibitem[{{Pavlov} \& {Gnedin}(1984)}]{PG:84} 
{Pavlov}, G.~G., \& {Gnedin}, Yu.~N.
Vacuum Polarization by a Magnetic Field and its Astrophysical Manifestations. 
Astrophys.\ Space Phys.\ Rev. 3, 197-253, 1984. 
 
\bibitem[{{Pavlov} {et~al.}(1994){Pavlov}, {Shibanov}, {Ventura} \& {Zavlin}}] 
 {Pavlovetal:94} {Pavlov}, G.G., {Shibanov}, Yu.A., {Ventura}, J. 
\& {Zavlin}, V.E. Model atmospheres and radiation of magnetic neutron stars: 
Anisotropic thermal emission. A\&A 289, 837-845, 1994.
 
\bibitem[{{Pavlov} {et~al.}(2004){Pavlov}, 
  {Sanwal}, {Garmire}, \& {et al.}}]{Pavlovetal:04} 
{Pavlov}, G.~G., {Sanwal}, D., {Teter}, M.A., Central Compact Objects in Supernova Remnants.
 In: 
Young Neutron Stars and Their Enviroments (Proceedings of the 
IAU Symp. 218), ed.\ F.~Camilo \& B.~M. Gaensler 
(ASP, San Francisco), 239-246, 2004.
 
\bibitem[{{Potekhin} \& {Chabrier}(2003)}]{PCh:03} 
{Potekhin}, A.Y. \& {Chabrier} G. 
Equation of State and Opacities for Hydrogen Atmospheres of Neutron Stars 
with Strong Magnetic Fields. ApJ 585, 955-974, 2003.
 
\bibitem[{{Potekhin} \& {Chabrier}(2004)}]{PCh:04} 
{Potekhin}, A.Y. \& {Chabrier} G. 
Equation of State and Opacities for Hydrogen Atmospheres of Magnetars.
ApJ 600, 317-323, 2004.
 
\bibitem[{{Potekhin} {et al.}(2004){Potekhin}, {Lai}, {Chabrier} \& {Ho}}] 
{Potekhinetal:04} 
{Potekhin}, A.Y., {Lai}, D., {Chabrier} G., \& {Ho}, W.~C.~G. 
Electromagnetic Polarization in Partially Ionized Plasmas with Strong Magnetic Fields 
and Neutron Star Atmosphere Models. ApJ 612, 1034-1043, 2004. 
 
\bibitem[{{Rajagopal} {et~al.}(1997){Rajagopal}, {Romani}, \& 
  {Miller}}]{Rajagopaletal:97} 
{Rajagopal}, M., {Romani}, R.~W., \& {Miller}, M.~C. 
Magnetized Iron Atmospheres for Neutron Stars.
ApJ 479, 347-356, 1997. 
 
\bibitem[{{Sanwal} {et~al.}(2002){Sanwal}, {Pavlov}, {Zavlin}, \& 
  {et~al.}}]{Sanwaletal:02} 
{Sanwal}, D., {Pavlov}, G.G., {Zavlin}, V.E. et. al. 
Discovery of Absorption Features in the X-Ray Spectrum of an Isolated Neutron Star.
ApJ 574, L61-L64, 2002.
 
\bibitem[{{Schwope} {et~al.}(2005){Schwope}, {Hambaryan}, {Haberl} 
 {et~al.}}]{Swopeetal:05} 
{Schwope}, A.D, {Hambaryan}, V., {Haberl}, F.  {et~al.} 
The pulsed X-ray light curves of the isolated neutron star RBS1223.
A\&A 441, 597-604, 2005. 
 
\bibitem[{{Schwope} {et~al.}(2007){Schwope}, {Hambaryan}, {Haberl} 
\& {Motch}}]{Swopeetal:07} 
{Schwope}, A.D, {Hambaryan}, V., {Haberl}, F. \& {Motch}, C. 
The complex X-ray spectrum of the isolated neutron star RBS1223.  A\&S 308, 619-623, 2007. 
 
\bibitem[{{Shibanov} {et~al.}(1992){Shibanov}, {Zavlin}, {Pavlov}, \& 
  {Ventura}}]{Shibanovetal:92} 
{Shibanov}, I.~A., {Zavlin}, V.~E., {Pavlov}, G.~G., \& {Ventura}, J. 
Model atmospheres and radiation of magnetic neutron stars. I - The fully ionized case.
 A\&A 266, 313-320, 1992. 
 
\bibitem[{{Suleimanov} \& {Werner}(2007)}]{Sul.Wer:07} 
{Suleimanov}, V. \& {Werner}, K. 
Importance of Compton scattering for radiation spectra of isolated neutron stars with 
weak magnetic fields.
A\&A 466, 661-666, 2007. 

\bibitem[{{Tiengo} \& {Mereghetti}(2007)}]{Tie.Meregh:07} 
{Tiengo}, A. \& {Mereghetti}, S. 
XMM-Newton Discovery of 7 s Pulsations in the Isolated Neutron Star RX J1856.5-3754.
ApJ 657, L101-L104, 2007. 
 
\bibitem[{{Tr\"umper} {et~al.}(2004) {Tr\"umper}, {Burwitz}, {Haberl} \& {Zavlin}}]{Trumperetal:04} 
 {Tr\"umper}, J.E., {Burwitz}, V., {Haberl}, F. \& {Zavlin}, V.E. 
The puzzles of RX J1856.5-3754: neutron star or quark star? 
Nuclear Phys. B Proc. Suppl. 132, 560-565, 2004. 
 
\bibitem[{{van Adelsberg} \& {Lai}(2006)}]{vAL:06} 
{van Adelsberg}, M. \& {Lai}, D. 
Atmosphere models of magnetized neutron stars: QED effects, radiation spectra 
and polarization signals.  MNRAS 373, 1495-1522, 2006. 
 
\bibitem[{{van Kerkwijk} \& {Kaplan}(2007)}]{vKK:07} 
{van Kerkwijk}, M.~H., \& {Kaplan}, D.~L. 
Isolated neutron stars: magnetic fields, distances,and spectra. A\&SS 308, 191-201, 2007. 
 
\bibitem[{{Zavlin} {et al.}(1995)}]{Zavlin:95} 
{Zavlin}, V.~E., Pavlov, G. G., Shibanov, Y. A., \& Ventura, J.
Thermal radiation from rotating neutron star: effect of the magnetic field 
and surface temperature distribution. A\&A 297, 441-450, 1995. 

\bibitem[{{Zavlin}(2009)}]{Zavlin:09} 
{Zavlin}, V.~E. Theory of radiative transfer 
in neutron star atmospheres and its applications.
In: Neutron Stars and Pulsars (Proceedings of the 
  363.\ WE-Heraeus Seminar),   
  ed.\ W.~Becker (Springer, New York), 181-211, 2009. 
 

\end{thebibliography}
\end{document}